\documentstyle[12pt,amssymb,epsf,epsfig,wrapfig]{article}
\begin{document}

\begin{center}
{\bfseries TRANSVERSITY AND THE POLARIZED\\ DRELL-YAN PROCESS IN $p\bar{p}%
\rightarrow \mu ^{+}\mu ^{-}X$}
\renewcommand{\thefootnote}{\fnsymbol{footnote}}
\footnote[1]{\it Presented at the Spin-05 - XI Workshop on high energy 
spin physics, Dubna, Sep. 27-Oct. 1, 2005}

\renewcommand{\thefootnote}{\arabic{footnote}}
\vskip 5mm
A.V. Efremov$^{1}$, O.V. Teryaev$^{1}$ and \underline{P. Z\'{a}vada}$^{2}$

\vskip 5mm
{\small
(1) {\it
Bogoliubov Laboratory of Theoretical Physics, JINR, 141980 Dubna, Russia
}\\
(2) {\it
Institute of Physics, Academy of Sciences of the Czech Republic, Na Slovance 2, CZ-182 21 Prague 8
}}
\end{center}

\vskip 5mm
\begin{abstract}
Estimates are given for the double spin asymmetry in lepton-pair production
from collisions of transversely polarized protons and antiprotons for the
kinematics of the recently proposed PAX experiment at GSI on the basis of
predictions for the transversity distribution from the probabilistic
quark-parton model developed earlier.
\end{abstract}

\vskip 8mm
\section{Introduction}

The leading structures of the nucleon in deeply inelastic scattering
processes are described in terms of three twist-2 parton distribution
functions -- the unpolarized $f_1^a(x)$, helicity $g_1^a(x)$, and
transversity $h_1^a(x)$ distribution. Owing to its chirally odd nature $%
h_1^a(x)$ escapes measurement in deeply inelastic scattering experiments
which are the main source of information on the chirally even $f_1^a(x)$ and 
$g_1^a(x)$. The transversity distribution function was originally introduced
in the description of the process of dimuon production in high energy
collisions of transversely polarized protons \cite{Ralston:ys}.

Alternative processes have been discussed. Let us mention here the Collins
effect \cite{Collins:1992kk} which, in principle, allows to access $%
h_{1}^{a}(x)$ in connection with a fragmentation function describing a
possible spin dependence of the fragmentation process, see also \cite%
{Mulders:1995dh} and references therein. Recent and/or future data from
semi-inclusive deeply inelastic scattering (SIDIS) experiments at HERMES %
\cite{Airapetian:1999tv}, CLAS \cite{Avakian:2003pk} and COMPASS \cite%
{LeGoff:qn} could be (partly) understood in terms of this effect \cite%
{DeSanctis:2000fh,Ma:2002ns,Efremov:2001cz}. Other processes to access $%
h_{1}^{a}(x)$ have been suggested as well, see the review \cite%
{Barone:2001sp}. However, in all these processes $h_{1}^{a}(x)$ enters in
connection with some unknown fragmentation function. Moreover these
processes involve the introduction of transverse parton momenta, and for
none of them a strict factorization theorem could be formulated so far. The
Drell-Yan process remains up to now the theoretically cleanest and safest
way to access $h_{1}^{a}(x)$.

The first attempt to study $h_1^a(x)$ by means of the Drell-Yan process is
planned at RHIC \cite{Bland:2002sd}. Dedicated estimates, however, indicate
that at RHIC the access of $h_1^a(x)$ by means of the Drell-Yan process is
very difficult \cite{Bunce:2000uv,Bourrely:1994sc}. This is partly due to
the kinematics of the experiment. The main reason, however, is that the
observable double spin asymmetry $A_{TT}$ is proportional to a product of
transversity quark and antiquark distributions. The latter are small, even
if they were as large as to saturate the Soffer inequality \cite%
{Soffer:1994ww} which puts a bound on $h_1^a(x)$ in terms of the better
known $f_1^a(x)$ and $g_1^a(x)$.

This problem can be circumvented by using an antiproton beam instead of a
proton beam. Then $A_{TT}$ is proportional to a product of transversity
quark distributions from the proton and transversity antiquark distributions
from the antiproton (which are connected by charge conjugation). Thus in
this case $A_{TT}$ is due to valence quark distributions, and one can expect
sizeable counting rates. The challenging program how to polarize an
antiproton beam has been recently suggested in the {\bf P}olarized {\bf A}%
ntiproton e{\bf X}periment (PAX) at GSI \cite{PAX}. The technically
realizable polarization of the antiproton beam of about $(5-10)\%$ and the
large counting rates -- due to the use of antiprotons -- make the program
promising.

In this note we shall make quantitative estimates for the Drell-Yan double
spin asymmetry $A_{TT}$ in the kinematics of the PAX experiment. For that we
shall stick to the description of the process at LO QCD. NLO corrections for 
$A_{TT}$ have been shown to be small \cite{Bunce:2000uv,NLO,Shimizu:2005fp}.
Similar estimations were done earlier \cite{Anselmino:2004ki,Efremov:2004qs}
using different models for the transversity distribution. Here for
transversity distribution we shall use the result of the covariant
probabilistic model developed earlier \cite{zav,tra}. In this model
the quarks are represented by quasifree fermions on mass shell and their
intrinsic motion, which has spherical symmetry and is related to the orbital
momentum, is consistently taken into account. It was shown, that the model
nicely reproduces some well-known sum rules. The calculation was done from
the input on unpolarized valence quark distributions $q_{V}$ and it was
shown, that assuming $SU(6)$ symmetry, a very good agreement with
experimental data on the proton spin structure functions $g_{1}$ and $g_{2}$
can be obtained.

\section{Transversity and the dilepton\\ transverse spin asymmetry}

In the paper \cite{tra} we discussed the transversity distribution in the
mentioned quark-parton (QPM) model. This model, in the limit of massless
quarks, implies the relation between the transversity and the corresponding
valence quark distribution: 
\begin{equation}
\delta q(x)=\varkappa \cos \eta _{q}\left( q_{V}(x)-x^{2}\int_{x}^{1}\frac{%
q_{V}(y)}{y^{3}}dy\right) .  \label{dy1}
\end{equation}%
The factors $\cos \eta _{q}$ represent relative contributions to the proton
spin from different quark flavors, which for the assumed $SU(6)$ symmetry
means, that $\cos \eta _{u}=2/3$ and $\cos \eta _{d}=-1/3$. The factor $%
\varkappa $\ depends on the way, in which the transversity is calculated:

{\it i)} Interference effects are attributed to the quark level only, then $%
\varkappa =1$. In this approach the relation between the transversity and
the usual polarized distribution is obtained%
\begin{equation}
\delta q(x)=\Delta q(x)+\int_{x}^{1}\frac{\Delta q(y)}{y}dy,  \label{dy2}
\end{equation}%
which means, that the resulting transversity distribution is roughly twice
as large as the usual $\Delta q$. The signs of both the distributions are
simply correlated. Soffer inequality in this approach is violated for the
case of large negative quark polarization, when $\cos \eta _{q}<-1/3$, which
means that the proton $d-$quarks in the $SU(6)$ scheme are just on the
threshold of violation.

{\it ii)} Interference effects at parton-hadron transition stage are
included in addition, but the result represents only upper bound for the
transversity. This bound is more strict than the Soffer one and roughly
speaking, our bound is more restrictive for quarks with\ a greater
proportion of intrinsic motion and/or smaller (or negative) portion in the
resulting polarization. No simple correspondence between the signs of actual
transversity and $\Delta q$ follows from this approach. In this scenario: $%
\varkappa =\cos ^{2}(\eta _{q}/2)/\cos \eta _{q}$.

Following the papers \cite{Anselmino:2004ki,Efremov:2004qs}, the
transversity can be measured from the Drell-Yan process $q\bar{q}\rightarrow
l^{+}l$ $^{-}$ in the transversely polarized $p\bar{p}$ collisions in the
proposed PAX experiment. The transversity can be extracted from the double
transverse spin asymmetry%
\begin{equation}
A_{TT}(y,Q^{2})=\frac{\sum_{q}e_{q}^{2}\delta q(x_{1},Q^{2})\delta
q(x_{2},Q^{2})}{\sum_{q}e_{q}^{2}q(x_{1},Q^{2})q(x_{2},Q^{2})};\qquad
x_{1/2}=\sqrt{\frac{Q^{2}}{s}}\exp (\pm y),  \label{dy3}
\end{equation}%
where, using momenta $P_{1},P_{2}$ of the incoming proton$-$antiproton pair
and the momenta $k_{1},k_{2}$ of the outgoing lepton pair, one defines the
physical observables 
\begin{equation}
s=\left( P_{1}+P_{2}\right) ^{2},\qquad Q^{2}=\left( k_{1}+k_{2}\right)
^{2},\qquad y=\frac{1}{2}\ln \frac{P_{1}(k_{1}+k_{2})}{P_{2}(k_{1}+k_{2})}.
\label{dy4}
\end{equation}%
The variable $y$ can be interpreted as the rapidity of lepton pair. The
asymmetry $A_{TT}$ is obtained from the cross sections corresponding to the
different combinations of transverse polarizations in the incoming $p\bar{p}$
pair%
\begin{equation}
A_{TT}(y,Q^{2})=\frac{1}{\hat{a}_{TT}}\frac{d\sigma ^{\uparrow \uparrow
}-d\sigma ^{\uparrow \downarrow }}{d\sigma ^{\uparrow \uparrow }+d\sigma
^{\uparrow \downarrow }};\qquad \hat{a}_{TT}=\frac{\sin ^{2}\theta }{1+\cos
^{2}\theta }\cos (2\varphi ),  \label{dy5}
\end{equation}%

\begin{wrapfigure}[19]{RT!}{.5\textwidth} 
\begin{center}
\vspace{-18mm}
\epsfig{file=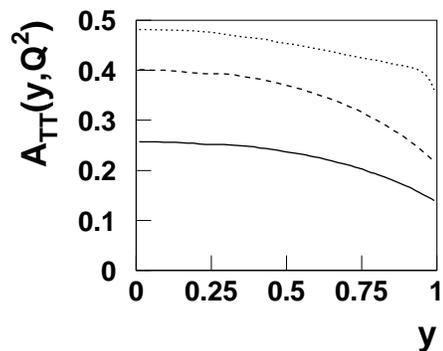, width=.5\textwidth}
\end{center}
\vspace{-6mm}
\caption{\small 
Double spin asymmetry at $Q^{2}=4GeV/c$ is calculated using two
transversity approaches: Interference effects are attributed to quark level
only {\it (solid line)}. Interference effects at parton-hadron transition
stage are included in addition {\it (dashed line)}, this curve represents
upper bound only. Dotted curve corresponds to the calculation based on
chiral quark-soliton model \cite{Efremov:2004qs}.}
\label{fi1}
\end{wrapfigure}
where the last expression corresponds to the double spin asymmetry in the
QED elementary process, $q\bar{q}\rightarrow l^{+}l^{-}$.
So using the above formulas, one can calculate the double spin asymmetry (%
\ref{dy3}) from the valence quark distribution according to the relation (%
\ref{dy1}). In Fig. \ref{fi1} the result of the calculation is shown.

The normalized input on the proton valence quark distribution was taken from
Ref. \cite{msr}, which corresponds to $Q^{2}=4GeV^{2}$ and the energy
squared of $p\bar{p}$ system is taken $45GeV^{2}$ in an accordance with the
assumed PAX kinematics. In the same figure the curve obtained at $%
Q^{2}=5GeV^{2}$ from the calculation \cite{Efremov:2004qs} based on the
chiral quark-soliton model \cite{bochum} is shown for a comparison. 
All curves in
this figure are based on the same parameterization \cite{grv} of the
distribution functions $q(x,Q^{2})$ appearing in the denominator in Eq. (\ref%
{dy3}). Obviously, our calculation gives a lower estimate of the $A_{TT}$
and one of possible reasons can be the effect of quark intrinsic motion,
which, as we have shown, can play role not only for the spin function $g_{1}$%
\cite{zav}, but also for the transversity $\delta q$ \cite{tra}. In an
accordance with \cite{Anselmino:2004ki},\ the motion of the lepton pair can
be described alternatively with the using the variable
\begin{equation}
x_{F}=\frac{2q_{L}}{\sqrt{s}}=x_{1}-x_{2}=2\sqrt{\frac{Q^{2}}{s}}\sinh y.
\label{dy6}
\end{equation}%
\begin{wrapfigure}[16]{RH}{.5\textwidth} 
\vspace{-15mm}
\begin{center}
\epsfig{file=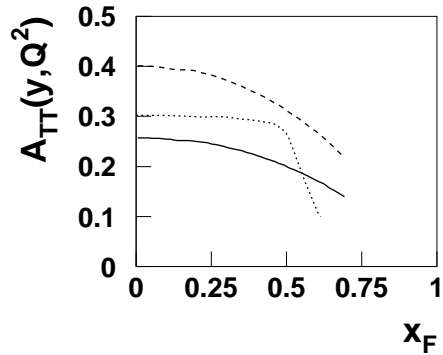, width=.5\textwidth}
\end{center}
\vspace{-8mm}
\caption{\small 
Double spin asymmetry: The solid and dashed lines are the same as in
the previous figure, but here their dependence on $x_F$ is displayed. Dotted
line is corresponding estimate from \protect\cite{Anselmino:2004ki}.}
\label{fi2}
\end{wrapfigure}

In the Fig. \ref{fi2} the estimation of asymmetry obtained in the cited
paper for $Q^{2}=4GeV^{2}$ and at $s=45GeV^{2}$ is compared with our curves
from Fig. \ref{fi1}, in which the variable $\ y$ is replaced by the $x_{F}$,
whereas both the variables are related by the transformation (\ref{dy6}).

Apparently, for $x_{F}\leq 0.5$ the curve from \cite{Anselmino:2004ki} is
quite compatible with our results.

So, in both the figures we have the set of curves resulting from different
assumptions and the experiment should decide, which one gives the best fit
to the data. How many events is necessary for discriminating among the
displayed curves? After integrating over angular variables one gets%
\begin{equation}
A_{TT}=\frac{n_{+}-n_{-}}{n_{+}+n_{-}},  \label{dy7}
\end{equation}%
then%
\begin{equation}
\Delta A_{TT}=\sqrt{\left( \frac{\partial A_{TT}}{\partial n_{+}}\Delta
n_{+}\right) ^{2}+\left( \frac{\partial A_{TT}}{\partial n_{-}}\Delta
n_{-}\right) ^{2}}=2\sqrt{\frac{n_{+}n_{-}}{n_{+}+n_{-}}},  \label{dy8}
\end{equation}%
which implies%
\begin{equation}
\Delta A_{TT}=\sqrt{\frac{1-A_{TT}^{2}}{N_{ev}}};\qquad N_{ev}=n_{+}+n_{-}.
\label{dy9}
\end{equation}%
So for approximate estimate of the statistical error we obtain%
\begin{equation}
\Delta A_{TT}\lesssim \frac{1}{\sqrt{N_{ev}}},  \label{dy10}
\end{equation}%
where $N_{ev}$ is number of events related to the bin or interval of $y$ or $%
x_{F}$ in which the curves are compared. For example, if one requires $%
\Delta A_{TT}\leq 1\%$, which is error allowing to separate the curves in
presented figures, then roughly $10^{4}$ should be the number of events in
the considered bin or interval. Of course, this estimation assumes full
polarization of the colliding proton and antiproton. Since the expected
polarization of antiprotons at the PAX will hardly be better than $(5-10)\%$%
, the minimum number of events will be correspondingly higher. 

\section{Summary}

The covariant probabilistic QPM, which takes into account intrinsic quark
motion, was applied to the calculation of transverse spin asymmetry of
dileptons produced in the $p\bar{p}$ collisions in the conditions, which are
expected for the recently proposed experiment PAX. This asymmetry is
directly related to the transversity distributions of quarks inside the
proton.\ In our asymmetry calculation the two approaches for the
transversity, which differ in accounting for the interference effects, were
applied. Our obtained results are compared with the prediction based on the
quark-soliton model. One can observe, that quite different approaches give
the similar results, but both our curves are lower than that obtained from
the quark-soliton model. Our results for $x_{F}\leq 0.5$ are also well
compatible with the recent estimate \cite{Anselmino:2004ki}.

\paragraph{Acknowledgement.}
This work is supported by Votruba-Blokhintsev Programm of JINR, A.E. and
O.T. are partially supported by grants RFBR 03-02-16816. Further, this work has 
been supported in part by the Academy of Sciences of the
Czech Republic under the project AV0-Z10100502.


\end{document}